\documentclass[%
 reprint,
nofootinbib,
 amsmath,amssymb,
 aps,
]{revtex4-2}

\usepackage{natbib}
\usepackage{color}
\usepackage{graphicx}
\usepackage{dcolumn}
\usepackage{bm}
\usepackage{hyperref}
\usepackage{soul}

\newcommand{\gray}{$\gamma$-ray }
\newcommand{\sv}{\langle \sigma v \rangle}
\newcommand{\lnl}{\log \mathcal{L}}
\newcommand{\Da}{{\rm D}}
\newcommand{\Ra}{{\rm R}}
\newcommand{\svul}{\langle \sigma v \rangle_{\rm UL}}

\newcommand{\ajnold}[1]{\textcolor{black}{#1}}
\newcommand{\Dold}[1]{\textcolor{black}{#1}}
\newcommand{\reDold}[1]{\textcolor{black}{#1}}
\newcommand{\omold}[1]{\textcolor{black}{#1}}

\begin{document}

\preprint{APS/123-QED}

\title{
\omold{Dark matter constraints with stacked gamma rays scales with the number of galaxies}}

\author{Daiki Hashimoto}
 \email{hashimoto.daiki@f.mbox.nagoya-u.ac.jp}
\affiliation{%
 Division of particle and astrophysical sciences, Graduate School of Science, Nagoya University, Furocho Chikusa, Nagoya, 464-8602, Aichi, Japan
}%

\author{Atsushi J. Nishizawa}%
\affiliation{
 Division of particle and astrophysical sciences, Graduate School of Science, Nagoya University, Furocho Chikusa, Nagoya, 464-8602, Aichi, Japan\\
 Institute for Advanced Research, Nagoya University, Furocho Chikusa, Nagoya, 464-8602, Aichi, Japan
}%

\author{Masahiro Takada}%
\affiliation{%
 Kavli Institute for the Physics and Mathematics of the Universe (WPI), The University of Tokyo Institutes for Advanced Study (UTIAS),\\~~~The University of Tokyo, 5-1-5 Kashiwanoha, Kashiwa-shi, Chiba, 277-8583, Japan
}%

\author{Oscar Macias}%
\affiliation{%
 Kavli Institute for the Physics and Mathematics of the Universe (WPI), University of Tokyo, Kashiwanoha, Chiba, 277-8583, Japan\\
 GRAPPA Institute, Institute of Physics, University of Amsterdam, 1098 XH Amsterdam, The Netherlands
}%


\date{\today}

\begin{abstract}
\omold{Low-surface-brightness galaxies (LSBGs) are interesting targets for searches of dark matter emission due to their low baryonic content. However, predicting their expected dark matter emissivities is difficult because of observational challenges in their distance measurements. Here we present a stacking method that makes use of catalogs of LSBGs and maps of unresolved gamma-ray emission measured by the \textit{Fermi} Gamma-Ray Space Telescope. We show that, for relatively large number of LSBGs, individual distance measurements to the LSBGs are not necessary, instead the overall distance distribution of the population is sufficient in order to impose dark matter constraints. Further, we demonstrate that the effect of the covariance between two galaxies located closely---at an angular distance comparable to the size of the \textit{Fermi} point spread function---is negligibly small. As a case in point, we apply our pipeline to a sample of $\sim$800 faint LSBGs discovered by Hyper Suprime-Cam and find that, the 95 per cent confidence level upper limits on the dark matter annihilation cross-section scales with inverse of the number LSBGs. In light of this linear dependence with the number of objects, we argue this methodology could provide extremely powerful limits if it is applied to the more than $10^5$ LSBGs readily available with the Legacy Survey of Space and Time.}
\end{abstract}

\maketitle

\section{Introduction}

\omold{R}evealing the nature of dark matter (DM) 
\omold{is one of the} major challenges in modern cosmology and particle physics.
As one of the 
\omold{best} theoretical\omold{ly} motivated candidates for 
\omold{DM}, weakly interacting massive particles~(WIMPs) have been considered \citep{1996PhR...267..195J} and probed via different approaches: colliders, underground experiments, and astrophysical observations  \citep[e.g.][]{2005PhR...405..279B, 2018EPJC...78..203A}.
In the context of astrophysics, researchers often focus on probing signals from the self-annihilation or decay of WIMPs.
In the early universe, WIMPs are considered to be produced in thermal equilibrium with standard model~(SM) particles.
As these interact with each other, WIMPs can annihilate into SM particles, particularly heavy SM particles, such as $b\bar{b}$ and $W^+W^{-}$~\citep{2018RPPh...81f6201R, 2019arXiv190407915L}.
As assumed that WIMPs are of DM abundance in the Universe, WIMP were reported to have an annihilation cross-section of $\sim 2\times 10^{-26}\; \rm cm^3/s$,
which is known as the thermal relic cross-section \citep{PhysRevD.86.023506}.
As a result of the annihilation process, $\gamma$ rays can be produced directly or in secondary 
\omold{processes in which case more} massive states decay into 
\omold{lighter and stable ones}, particularly photons, electrons, positrons and neutrinos. Therefore, probing these particles induced by DM annihilation gives clues to specify DM properties.

The Fermi Large Area Telescope~(LAT) has revealed the \gray sky 
in the energy range from \omold{about} 20~MeV to \omold{approximately} 1~TeV. 
In the fourth catalog of 
\gray point sources~\citep{4FGL}, \omold{the Fermi team presented} the most detailed galactic diffuse emission 
\omold{model as well as } 
\omold{a} catalog containing about 5,000 objects with above $4\sigma$ significance detection. 
\omold{Out of all these,} more than 3,700 sources have been found to 
\omold{be} associated 
\omold{with} pulsars, supernova remnants and/or blazars. \omold{Interestingly, \textit{Fermi}-LAT observations have also been used to probe DM from different sky regions such as; the Galactic Center~\citep{2013PhRvD..88h3521G, 2014PhRvD..90b3526A, 2015JCAP...03..038C, 2016PDU....12....1D, 2017ApJ...840...43A}, Milky Way dwarf spheroidals~(MW dSphs)~\Dold{\citep{2015PhRvL.115w1301A, 2017ApJ...834..110A, 2018PhRvD..98h3008G, 2018arXiv181206986H, 2018PhRvD..97i5031B, 2020JCAP...02..012H}}, DM subhaloes \citep{2014GrCo...20...47B}, and the unresolved \gray background~\Dold{\citep{2013PhRvD..87l3539A, Fornasa:2015qua}} (UGRB), to mention just a few.}
\Dold{Of these targets, MW dSphs have provided the most robust DM constraint because the astrophysical background uncertainties in these objects is expected to be relatively small.}

\omold{The UGRB} is the residual emission obtained by subtracting the resolved point sources and galactic diffuse emission from the \gray observations.
\omold{As aforementioned, this component} 
has been \omold{extensively} studied to probe DM annihilation signals from e.g., low-redshift 
\omold{galaxy} catalogs, 
galaxy groups and galaxy clusters \citep{2014PhRvD..90b3514A, 2015ApJ...799...86A, 2015ApJ...800L..27A, 2015ApJ...807...77K, 2015ApJS..221...29C, 2015PhRvL.114x1301R, 2016PhRvD..94f3522S, 2017MNRAS.467.2706T, 2018PhRvL.120j1101L, 2018PhRvD..97f3005L, 2018PhRvL.121x1101A, 2019MNRAS.484.5256H, 2021MNRAS.501.4238B, 2021MNRAS.502.4039T}.
%
In our previous work~\citep{2020JCAP...01..059H}, \omold{we proposed to use a} low-surface-brightness galaxy~(LSBG), which has less than $\sim$23 $\rm mag/arcmin^2$ mean surface brightness, 
as a new target to probe DM annihilation signals in the UGRB.
\omold{In particular, }LSBGs are known to be highly dominated by DM \citep{2017MNRAS.470.1512W, 2019MNRAS.483.1754D} and have more massive halos than Milky Way dSphs ($\sim10^9 M_\odot$).
Furthemore, they have 
\omold{lower} star-formation activity, \gray \omold{contamination} from pulsars and supernova remnants, than ordinary galaxies or galaxy clusters \citep{2019MNRAS.484.4865P}.
However, owing to their faint fluxes, it is more difficult to measure precisely their redshifts than for ordinary galaxies. \omold{This explains why}, in our previous \omold{study}, \omold{we only used eight out of the nearly 800 LSBGs found with the Hyper Suprime-Cam~(HSC)~\cite{2018ApJ...857..104G}.} 

In this article, we propose a method \omold{to probe DM model parameters by using} a large number of objects 
\omold{with unknown} redshifts. 
\Dold{Instead of individual redshifts, we focus on the redshift distribution of an entire catalog, which is obtained by the clustering redshift method. 
Each object's redshift is randomly assigned from the distribution in order to model the \gray flux induced by the DM annihilation.  Using a composite likelihood analysis of the model fluxes and the UGRB data, we are able to compute constraints on the DM annihilation cross-section. In addition, we perform several validation checks of our methods using the full sample of the HSC-LSBG catalog and UGRB data from the \textit{Fermi}-LAT observation.}
%
\omold{Since running a maximum-likelihood analysis with so many objects is very computationally intensive, we initially assume that the $\gamma$-ray emission from each LSBG is independent of each other. We verified the validity of this assumption by estimating the covariance of the best-fit model parameters of neighboring LSBGs.}
\omold{Our main analysis method is that of a stacking analysis with the population of LSBGs. We discuss the power of this technique for the problem at hand. } 
Because LSBGs are potentially abundant in the local universe, and \omold{the upcoming} 
Legacy Survey of Space and Time~(LSST) \omold{is expected to discover many more such systems,} 
\omold{we anticipate that LSBGs can become competitive targets for indirect DM detection efforts. }

This paper is organized as follows.
In Section~\ref{sec:mth}, we 
\omold{present} the \omold{clustering} \Dold{redshift method for modeling $\gamma$-ray emissivities of LSBGs.}
In Section~\ref{sec:result}, we 
\omold{perform a stacking analysis using co-added unresolved $\gamma$-ray maps } 
\omold{and our sample of} HSC-LSBGs. 
\omold{In Section~\ref{sec:discussion}, }we present a scaling relation \omold{between the 95\% confidence level upper limit on the DM cross-section and the number of LSBGs.} 
Finally, we conclude 
in Section~\ref{sec:summary}.

\section{Theoretical framework}
\label{sec:mth}

In this section, we present our \omold{methodology} to constrain the DM annihilation cross-section 
\Dold{using the}
diffuse \gray background and massive nearby objects \omold{having no individual distance estimates. Note that our methodology is generic and could be applied to highly DM dense regions \omold{that would appear in the {\it Fermi} data as point sources} such as; dSphs, faint galaxies, and normal galaxies.} 
\omold{As discussed in the introduction, since we do not have individual distance measurements for our targets,} we require an overall probabilistic distribution of their \omold{positions} along the line of sight. \omold{In addition}, we assume that all objects are statistically independent, which dramatically simplifies the stacking likelihood analysis.
In Subsection~\ref{ssec:cov}, we evaluate the validity of this assumption.

\subsection{\omold{Predicted} \gray flux \omold{from} DM annihilations}
\label{ssec:model}
Through the DM annihilation process, \gray photons are generated directly or in the cascade process in which various final states (e.g. $b\bar{b}$, $W^+W^-$ and $\mu^+ \mu^-$) decay into more stable particles. The \gray flux ${d\Phi_{\rm ann}}/{dE}$ produced by DM annihilation can be modeled as follows,
\begin{equation}
\frac{d\Phi_{\rm ann}}{dE} = J \times \frac{\langle \sigma v \rangle}{8\pi m_{\chi}^{2}}
\sum_{i}{\rm Br}_{i}\frac{dN_{i}}{dE},
\label{eq:ann_flux}
\end{equation}
where $m_\chi$ is the DM mass and $\sv$ is the \omold{velocity-averaged DM annihilation cross-section}. 
\omold{The parameters }
${\rm Br}_i$ and $\frac{dN_i}{dE}$ are the branching ratio and \gray energy spectrum 
in the $i-$th annihilation channel, respectively. 
In the analysis, we consider $b\bar{b}$ as a representative annihilation channel and obtain $dN_{b\bar{b}}/dE$ 
\omold{using the} \texttt{DMFIT}\footnote{\url{https://fermi.gsfc.nasa.gov/ssc/data/analysis/scitools/gammamc_dif.dat}}~\citep{Jeltema:2008hf} \omold{package} included in the official \texttt{fermipy} analysis software. 
\omold{The parameter} $J$ is the so-called J-factor, which is given by the DM halo properties as follows,
\begin{equation}
J = [1 + b_{\rm sh}(M_{\rm halo})] \int_{s} ds' \int_{\Omega} d\Omega' \rho^{2}_{\rm DM}(s',\Omega'),
\label{eq:j_factor}
\end{equation}
where $s'$, $\Omega'$ and $M_{\rm halo}$ are the line-of-sight vector, solid angle and halo mass of target objects, respectively. 
The annihilation signal may increase if we have clumpy substructures within a halo instead of the smooth halo. 
This can be effectively modeled as a boost factor, $b_{\rm sh}$ \citep{Hiroshima+2018}. For clarity, we set $b_{\rm sh}=1$ for all 
halos throughout this paper.
The DM density profile is assumed to follow the Navarro-Frenk-White (NFW) shape \citep{NFW},
\begin{equation}
\rho_{\rm DM}(r) \propto \frac{1}{c r/r_{\rm vir} \left[(c r/r_{\rm vir}) + 1\right]^{2}},
\end{equation}
where $c(M_{\rm halo})$ is the concentration parameter. 
The concentration parameter primarily depends on the halo mass and we compute it using the fitting formula to model the scaling relation of $c(M)$ with halo mass, calibrated using the high-resolution N-body simulations \citep{2019ApJ...871..168D}.
We convert the halo mass to the concentration by using the \texttt{COLOSSUS} package~\citep{2018ApJS..239...35D}.

The full description of J-factor estimation can be found in \cite{2020JCAP...01..059H}. A brief summary is as follows.
We first convert the observed magnitude into the absolute magnitude. The V-band apparent magnitude can be converted from the $gri$ system as $V=g-0.59(g-r)-0.01$ \citep{2005AJ....130..873J}. Now, the situation is that we do not have the distance to individual galaxies but an overall distribution.
However, we can formally write the distance of each galaxy as a random draw from the distribution: $d \in dN/dz(z)$. For the particular realization of the random draw of the distance with the binding condition of $\langle d \rangle = dN/dz$, the absolute magnitude can be computed; thus, the halo mass can be derived by assuming the relations of mass-to-light ratio \citep{2008MNRAS.390.1453W} and stellar to halo mass ratio \citep{2013MNRAS.428.3121M}.

Given that we have the overall $dN/dz$ distribution, with measurement errors, we can simulate the effect of neglecting distance to individual galaxies. Figure~\ref{fig:sumJ} shows the distribution of the total J-factor after stacking $N_{\rm st}$ galaxies. For this plot, we use the HSC-Y1 LSBG sample, where no distance is available. The $N_{\rm st}$ LSBGs are randomly selected from the parent sample 500 times. The scatter in the figure corresponds to the 95\% range due to the effect of the sample variance.

\subsection{Measurement of the redshift distribution of photometric samples}
\label{ssec:dndz_method}
In this subsection, we describe a method to estimate the $dN/dz$ distribution from spatial clustering; \omold{the so-called} clustering redshift method. 
The application of this method to the data is shown in Subsection~\ref{ssec:dndz_measurement}.

Given two galaxy samples in an overlapped region, even if the galaxies in the two galaxy samples are statistically different, they both correlate with the underlying DM distribution. In the case where one galaxy sample has known redshifts and the other \omold{does not},
\omold{it is possible to take the} 
cross-correlation between the two galaxy samples \omold{to obtain a} 
statistical estimate of the redshift distribution of the galaxies with unknown redshifts.
We denote the galaxy samples with known and unknown redshifts as `\textit{spectroscopic (spec-z) sample}' and `\textit{photometric sample}', respectively. The angular cross-correlation function can be factorized as follows \citep{2008ApJ...684...88N},
\begin{equation}
    w(\theta) = \int_0^{\infty} dz \frac{dN_p}{dz}\frac{dN_s}{dz}b_p(z)b_s(z)w_{\rm DM}(z,\theta),
    \label{eq:w_theta}
\end{equation}
where subscripts $s$ and $p$ represent spec-z and photometric samples, respectively.
$dN/dz$ represents the redshift distribution normalized to unity,
$b(z)$ is a linear bias, and $w_{\rm DM}(z,\theta)$ is the angular correlation function of DM. 
Note that the mass of the employed DM model is high enough \omold{so as} not to affect the DM clustering pattern itself.
We define an integrated cross-correlation $\bar{w}$ with a weighting function $W(\theta)$ as 
\begin{equation}
    \bar{w} = \int_{\theta_{\rm min}}^{\theta_{\rm max}} d\theta~ W(\theta) w(\theta),
\label{eq:norm_w}
\end{equation}
where the weight $W$ is introduced so that the signal-to-noise ratio of $\bar{w}$ can be optimized. Following \citep{2013arXiv1303.4722M}, we empirically adopt $W=\theta^{-1}$.
In \omold{practice},
we divide the spec-z sample into narrow redshift bins so that $dN_s/dz$ can be approximated by the narrow top-hat function; $dN_s^i/dz \simeq 1/\Delta z$ if $z_i<z<z_{i+1}$.
\omold{After this,}
we rewrite Equation~\ref{eq:norm_w} at $z=z_i$ as 
\begin{equation}
\bar{w}(z_i) \approx \frac{dN_p}{dz}(z_i) b_{\rm p}(z_i)b_{\rm s}(z_i)\bar{w}_{\rm DM}(z_i),
\label{eq:dndz}
\end{equation}
%
where $\bar{w}_{\rm DM}(z)$ can be defined similarly to Equation~\ref{eq:norm_w} by replacing $w(\theta)$ with $w_{\rm DM}(z, \theta)$.

\Dold{Here, we assume that the biases of the LSBG and spec-z sample are constant over redshift because the redshift range that we consider is narrow. Although $\bar{w}_{\rm DM}(z)$ is fully derived from the standard cold DM theory including the non-linear matter clustering evolution, we do not need to estimate the absolute value of it. What we need to consider on $\bar{w}_{\rm DM}$ is the redshift evolution. Given the narrow redshift range, in our analysis, we simply replace $\bar{w}_{\rm DM}(z)$ with the square of the linear growth factor, $D^2(z)$.}

\section{Data} 
\label{sec:data}

\omold{We start by validating our methodology. For this,} we apply \omold{the aforementioned pipeline} to the LSBG samples \citep{2018ApJ...857..104G} observed with the HSC survey. The 781 LSBG samples in $\sim 200$ deg$^2$ of the HSC region do not have individual distance measure, but we have the spec-$z$ samples from NASA Sloan Atlas~(NSA) \citep{2011AJ....142...31B} on an almost fully overlapped sky area. 
\omold{Lastly, we use the \textit{Fermi}-LAT unresolved extragalactic} 
\gray background 
by applying the composite likelihood analysis \omold{to the sample of HSC-LSBGs.  } 
\omold{Below we provide details of the datasets used in our study.}

\subsection{HSC-LSBG catalog}
\label{ssec:lsbg}
HSC is a wide-field camera, attached to the prime focus of the Subaru telescope, covering $\sim$$1.5^{\circ}$ diameter field of view with 0.17 arcsec pixel scale \citep{2018PASJ...70S...1M, 2018PASJ...70S...2K}. 
The HSC Subaru Strategic Program survey consists of three layers, wide, deep and ultradeep layers; the wide layer has five broad photometric bands $g, r, i, z$ and $y$. 
As described in a report of the second data release of the survey by \cite{2019PASJ...71..114A}, the wide layer has a depth of $24.5$--$26.6$ in the 5 filters for 5 $\sigma$ point-source detection. 
In the final data release, the survey will cover 1400 $\rm deg^2$ sky in a depth of $i \sim$26 mag. 

Since the HSC pipeline, \texttt{hscpipe}, is not optimized for detecting and measuring diffuse objects, HSC images are reduced first by the \texttt{hscpipe} and then \texttt{SExtractor} is used for detection and measurements.
\cite{2018ApJ...857..104G} processed an HSC dataset with three broad bands ($g,r$ and $i$) on a patch-by-patch basis over $\sim 200~\rm deg^2$ and produced a catalog comprising 781 LSB objects, which have a mean surface brightness in $g-$band $>24.3 \rm mag/arcsec^2$.
Briefly, the following processes have been executed: 
(i) bright sources and associated diffuse lights are subtracted from images to avoid contamination to LSB objects detection;
(ii) after the Gaussian smoothing with full depth at half maximum of $1''$, sources with the half-light radius $r_{1/2}$ satisfying $2.5'' < r_{1/2} < 20''$ are extracted; 
(iii) the sources are selected by applying reasonable color cuts to remove optical artifacts and distant galaxies; 
(iv) by modeling the surface brightness profiles of LSBG candidates, astronomical false positive are removed; 
(v) by visual inspection, false candidates such as point-sources with diffuse background lights are removed and then 781 LSBGs are finally left.

To minimize a possible contamination of astrophysical $\gamma$-ray photons,it might be useful to restrict our analysis to quiescent galaxies, because such galaxies do not have ongoing star-formation activities and therefore unlikely contain high-energy astrophysical sources such as supernova remnants and AGNs, at least compared to star-forming blue galaxies.
We divide the sample into red and blue LSBGs, where red is defined by $g-i \geq 0.64$ and blue $g-i < 0.64$, which include 450 and 331 objects, respectively. This color selection roughly corresponds to the galaxy age of 1 Gyr for a 0.4 $\times$ solar metallicity galaxy \citep{2018ApJ...857..104G}.
\Dold{In Figure~\ref{fig:NSA}, we show the sky distribution of our LSBG sample in green dots.}
For a random catalog corresponding to the LSBG catalog, we employ the random catalog of the HSC photometric data, and randomly resample it such that the number density is roughly 10 times larger than the LSBG density. We also apply the bright star mask to the random catalog.

\subsection{NSA sample}
\label{ssec:nsa}

\begin{figure}
 \begin{center}
  \hspace{-0.3cm}
  \includegraphics[width=8cm]{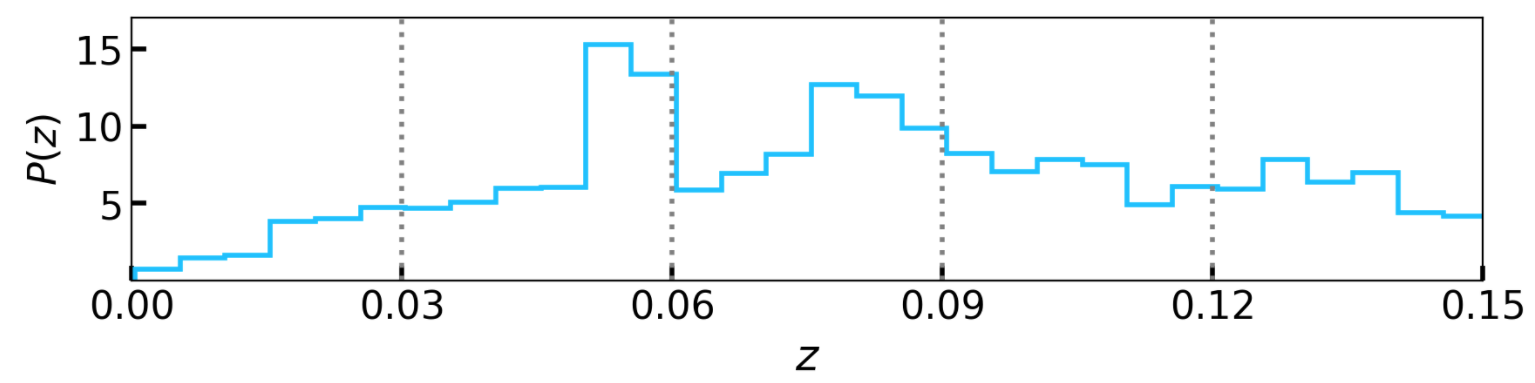}\\
  \hspace{-0.5cm}
  \includegraphics[width=9cm]{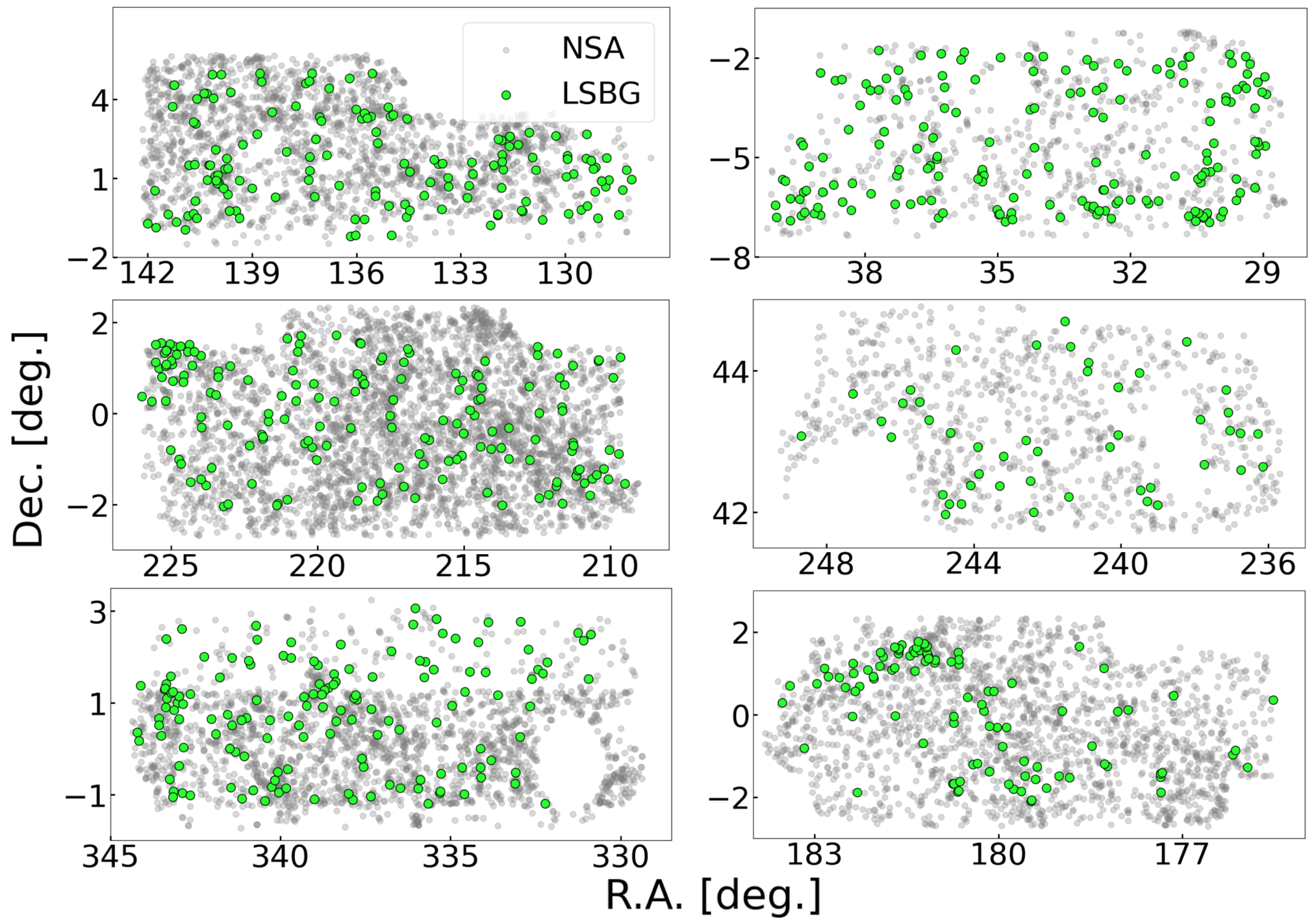}
 \end{center}
 \caption{
 (\Dold{\textit{Top}) Redshift distribution of the NSA sample.
 The vertical lines show boundaries of redshift bins for cross correlation analysis.
 (\textit{Bottom}) Sky distributions of the HSC-LSBG sample (green dots) and NSA sample (gray dots).} 
 }
\label{fig:NSA}
\end{figure}

\omold{To measure the} $dN/dz$ distribution, we need reference spec-$z$ samples in the overlapped region in both sky-coverage and redshift range.
\Dold{Since the LSBGs have been detected in the local universe by radio observations~\citep{2019MNRAS.483.1754D, 2019ApJS..242...11L}, optical selected LSBGs are also expected to be in the local universe; thus, we need a low-redshift spec-$z$ sample.}
The NSA sample \footnote{\url{https://data.sdss.org/sas/dr13/sdss/atlas/v1/nsa_v1_0_1.fits}} is a spec-z sample obtained from the spec-$z$ campaign of the Sloan Digital Sky Survey with the Galaxy Evolution Explorer data for the energy spectrum of the ultraviolet wavelength and includes objects up to $z = 0.15$ \Dold{including 11,820 objects in the overlap region of the LSBG. We show the sky position and redshift distribution in Figure \ref{fig:NSA}.}
Since the uniformity of the NSA sample is not guaranteed, we attempt to mitigate the non-uniformity as follows. First, we remove the sample from both HSC and NSA in the bright star masked regions.
We checked that, after removing the masked regions, the local number density of NSA galaxies smoothed with each HSC patch has a uniform distribution in most areas of the HSC regions we are working on, except for the low-density region in the VIMOS-VLT Deep Survey (Dec.$>$1).
We generate random catalogs including about 10 times more objects than that of the NASA galaxies.
In addition, we removed the edge regions in the HSC survey footprint for safety, because the exact survey window near the boundaries is difficult to define.

\subsection{\Dold{UGRB from LAT data}}
\label{ssec:ugrb}

\subsubsection{\Dold{LAT data}}
\label{sssec:gamma}
We analyze 
%
%
the \textit{Fermi}-LAT \texttt{Pass 8} data obtained from 2008-08-04 to 2016-08-02,
which 
\omold{contains several upgrades to account for a much improved knowledge of the instrument response function~\citep{2013arXiv1303.3514A}.} 
As recommended by the Fermi collaboration 
\footnote{\url{https://fermi.gsfc.nasa.gov/ssc/data/analysis/documentation/Cicerone/Cicerone_Data_Exploration/Data_preparation.html}},
we select the photon event class {\tt P8R3 SOURCE}~\citep{2012ApJS..203....4A}, 
which is suitable for point-source analyses. 
In addition, we apply the following selections, {\tt DATA\_QUAL>0, LAT\_CONFIG==1} and \texttt{P8R3\_SOURCE\_V2}, as the corresponding filter expression and the instrument response function for the event class, respectively.
We use photon counts in the energy range 500~MeV to 500~GeV, \omold{where we split the data into} 24 logarithmically spaced energy bins, \omold{and spatial bins of size} $0.1^{\circ}\times0.1^{\circ}$.
\omold{We selected} the width of the energy bins 
\omold{such that those are larger than the} energy dispersion \omold{of the instrument at any given bin, but} we require 
spatial bins to be smaller than the LAT PSF at all 
\omold{energies}.
The lower energy scale is determined to balance two opposing effects. If we decrease the low-energy limit, photons around bright sources would leak due to the broadening of the PSF, but if we increase the low-energy limit, the photon count statistics decreases.

For the composite analysis described in Subsection~\ref{ssec:composite}, we select 29 patches whose centers are located in the HSC regions with the separation of at least $3^{\circ}$ from each other and individual patches having $10^{\circ} \times 10^{\circ}$ sky coverage. 
Moreover, to avoid contaminating the \gray photons produced by the Earth's atmosphere interaction with high energy cosmic rays, we exclude the photon data with zenith angles greater than $100^{\circ}$.
In the LAT data analysis, we use \texttt{fermipy} (v1.0.1) \footnote{\url{https://fermipy.readthedocs.io/en/latest/}} \citep{2017ICRC...35..824W}, which is an open-source software package based on the {\tt Fermi Science Tools (v2.0.8)}\footnote{\url{https://fermi.gsfc.nasa.gov/ssc/data/analysis/software/}}


\subsubsection{UGRB construction and putative flux}
\label{sssec:measured_flux}


\omold{Our procedure to reconstruct the UGRB observations and compute the likelihood profile for each LSBG object is the same one introduced in \cite{2015PhRvL.115w1301A}.}
First, 
we perform the maximum likelihood analysis in each patch to optimize the normalization and spectral parameters of 
all \omold{the standard 4FGL \citep{4FGL}} sources 
in the patch -- including the galactic diffuse emission, isotropic emission, and \omold{extended} sources.
\omold{Specifically,} we adopt the standard \omold{Galactic diffuse emission model} \texttt{(gll\_iem\_v07.fits)} and the isotropic emission template \texttt{(iso\_P8R3\_SOURCE\_V2\_v01.txt)}\footnote{\url{https://fermi.gsfc.nasa.gov/ssc/data/access/lat/BackgroundModels.html}}, respectively.
The isotropic template represents isotropic contributions from undetected extragalactic sources and the residual cosmic ray 
\omold{background}. 

\begin{figure}
 \begin{center}
  \hspace{-0.5cm}
  \includegraphics[width=9cm]{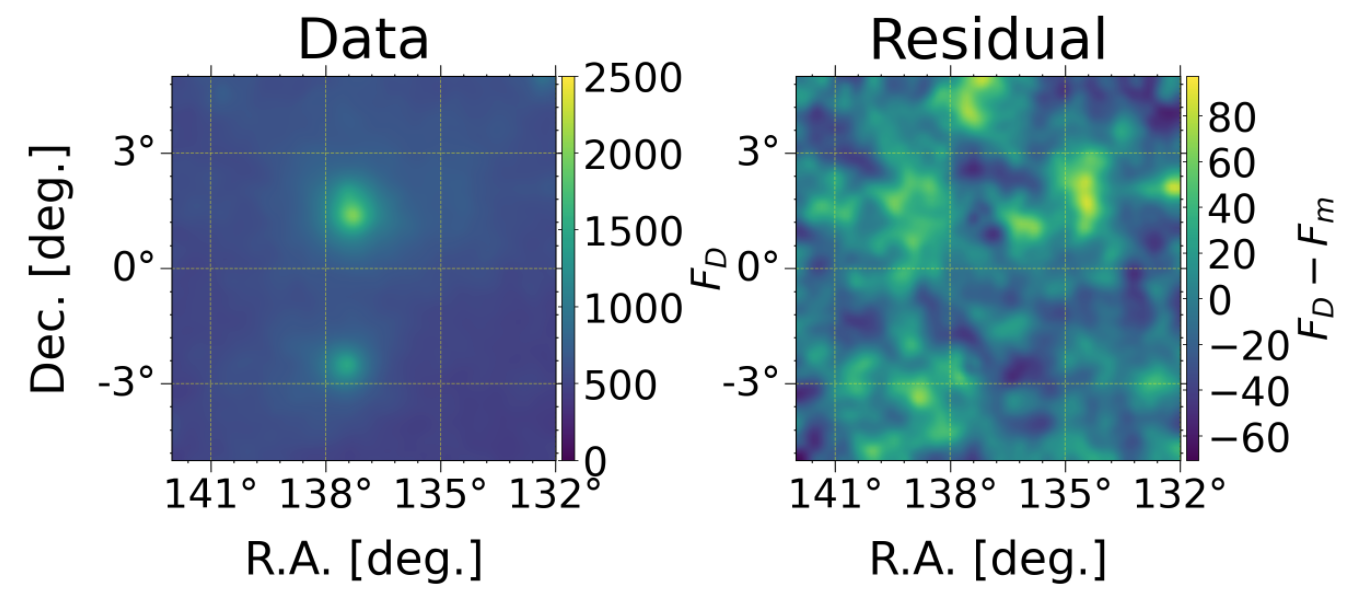}\\
  \hspace{-0.5cm}
  \includegraphics[width=8cm]{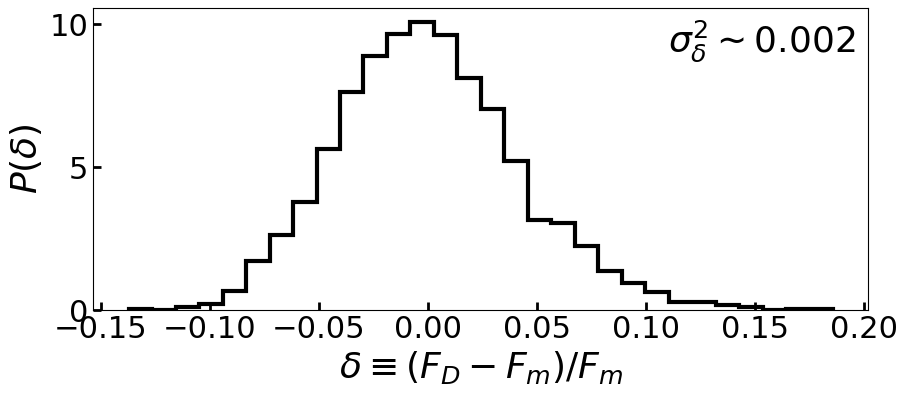}
 \end{center}
 \caption{\Dold{(\textit{Top}) An example of observed photon map (left) and model-subtracted residual map (right) in a single patch. The color bar shows the photon count.
 (\textit{Bottom}) Histogram of the relative fluctuation of the residual map.}
 }
\label{fig:gammaray_map}
\end{figure}

In Figure~\ref{fig:gammaray_map}, we show an example of the observed and residual counts 
\omold{for one of our} patches. 
\omold{In addition, we show a} histogram of the relative fluctuation of the residual counts map for the same patch. \omold{The} rms \omold{of the} fluctuation \st{being} \omold{is} $\sim 0.04$.
Note that we confirm \omold{that} 
\omold{our patches do not contain new} point sources with test statistics~(TS) larger than 25. 

\omold{Once reconstructed the UGRB,} we estimate the 
\omold{putative} flux 
\omold{of each} LSBG. 
According to the prescription described in the 2FGL catalog, the Bayesian method~\citep{1983NIMPR.212..319H} should be applied for the likelihood analysis of very faint sources \citep{2FGL}.
\Dold{We introduce free parameters $\alpha_{ij}$ representing the $i$-th point-source flux-amplitude 
in $j$-th energy bin.
}
Here we define TS as $TS \equiv -2\Delta \lnl$ and $\Delta \lnl$ is given by;
\begin{equation}
    \Delta \lnl = \lnl(\mathcal{D},\boldsymbol{\Theta}|\alpha_{i,j}=0) - \lnl (\mathcal{D},\boldsymbol{\Theta}|\alpha^{\rm max}_{i,j}),
    \label{eq:pl_flux}
\end{equation}
where $\mathcal{D}$ and $\boldsymbol{\Theta}$ are the LAT data and the nuisance parameters of all the \gray sources other than our target LSBGs. 
\omold{The best fit parameters}
$\alpha^{\rm max}_{i,j}$ \omold{are obtained } when $\lnl (\mathcal{D},\boldsymbol{\Theta}|\alpha_{i,j})$ is maximized. \omold{In the above equation} 
$\alpha_{i,j}=0$ means no target source.
We 
\omold{checked} that \omold{the} TS values of all our LSBGs are less than 1 for most energy bins, which 
\omold{justifies our choice of using} the Bayesian method 
\omold{introduced in} \cite{4FGL}.



\section{Result} 
\label{sec:result}

\subsection{Composite analysis}
\label{ssec:composite}
As described in Section~\ref{sssec:measured_flux}, because of the very low \gray signal of our LSBGs, \omold{we compute the 95\% C.L. upper limits on the DM annihilation cross-section parameter using the same approach as in~\citep{2020JCAP...01..059H}. } 
Since all likelihood values obtained at each object are assumed to be independent of each other, the composite likelihood ${\mathcal L}_{\rm st}$ for the full sample of targets can be expressed as follows;
\begin{equation}
 \log {\mathcal L}_{\rm st}(\alpha|\sv, J) = \sum_{i,j} \log {\mathcal L}^{\rm ann}_{i,j}(\alpha_{i,j}|\{ \sv, J_i\}),
\label{eq:stackL}
\end{equation}
where $J_i$ is the $J$-factor of the $i$-th target and the index $j$ runs over all energy bins. 
${\cal L}^{\rm ann}_{i,j}$ is the log-likelihood value for the $i$-th LSBG, which is obtained by the difference of the log-likelihood value from the one in the case of no source.

Now we calculate the J-factor for each object to evaluate the model flux described in Section~\ref{ssec:model}.
For the simplicity of Equation~\ref{eq:j_factor}, we need to consider the relationship between the PSF of the LAT instrument and the angular size of objects.
The PSF (68\% containment angles) decreases from $\sim 1.5^{\circ}$ to $\sim 0.1^{\circ}$ as \gray energy increases from 500~MeV to 500~GeV.
The angular size of LSBG is smaller than a few 10 arcsecs, hence smaller than the PSF in all energy bands.
Therefore, we can consider them as point-like sources in the likelihood computation.
Then, the integration of $\rho_{\rm DM}^2$ over the target volume in Equation~\ref{eq:j_factor} is reduced to;
\begin{equation}
    \int ds \int d\Omega \rho^2_{\rm DM}(s,\Omega)
    \rightarrow
    \int dV \rho^2_{\rm DM}(r) / d_A^2,
    \label{eq:rho2_int}
\end{equation}
where $d_A$ is the angular diameter distance to the object, which is given by assignment of distance randomly drawn followed the measured $dN/dz$ distribution.
Then, Equation~\ref{eq:rho2_int} is straightforwardly calculated; 
%
\begin{align}
 J &= (1 + b_{\rm sh}) \frac{M_{\rm halo}}{d_A^2}\frac{\Delta \rho_{c,z}c^{3}}{9} \times \\ \notag
   & ~~~~~~ \left[1 - \frac{1}{(1+c)^{3}}\right] \left[ \log(1+c) - \frac{c}{1+c} \right]^{-2},
\label{eq:finalJ}
\end{align}
\Dold{where $\Delta$ is the spherical overdensity set to be 200 and $\rho_{c, z}$ is the critical density at the redshift.}
Since our targets are regarded as point sources, our assumption is correct if the angular separations between objects are larger than the LAT PSF over all considered energy ranges; otherwise, it is incorrect because their mean surface number density is about 4 per $\rm deg^2$. 
We will further discuss the parameter correlation between neighboring objects in Section~\ref{ssec:cov}.
In our procedure, we assume that the LSBG flux is positive definite, which implies that the data are well-described by a $\chi^2/2$ distribution rather than $\chi^2$. 
As such, the 95\% C.L. upper limits on the cross section $\svul$ are given when the $\Delta\lnl_{\rm st}(\alpha_{i,j}|\sv,J) \sim -3.8/2$. 

\subsection{$dN/dz$ measurement}
\label{ssec:dndz_measurement}

Following the methodology described in Subsection~\ref{ssec:dndz_method}, we 
\omold{estimate} the $dN/dz$ distribution of the HSC-LSBG sample. \omold{For this,} we divide the reference redshift sample into five equally-separated bins from $z=0$ to $0.15$. 
Then, we 
\omold{compute} the cross correlation between the sample in each bin and the entire LSBG sample. 
The angular cross correlation is computed \omold{in the angular bin $0.1^{\circ}<\theta<1.0^{\circ}$, which is further divided in smaller logarithmic-scaled bins. We use the} estimator \citep{Landy-Szalay:1993}, 
%
\begin{equation}
   w(z_m,\theta_i)= 
   \frac{
   \Da_{p}\Da_{s,m} - 
   \Da_{p}\Ra_{s,m} - 
   \Ra_{p}\Da_{s,m} + 
   \Ra_{p}\Ra_{s,m}}
   {\Ra_{p}\Ra_{s,m}
   }, 
\end{equation}
where $\Da\Da$ or $\Da\Ra$ 
represent the normalized number of pairs separated within the $i-$th angular bin between data and data
or data and random, 
respectively \reDold{(in our case, ``data" and ``random" represent the spec-z or LSBG samples in our datasets and the corresponding random samples, respectively).} 
Subscripts $p,s$, and $m$ represent the photometric sample, and reference sample in the $m-$th redshift bin, respectively. 

\begin{figure}
 \begin{center}
  \includegraphics[width=8cm]{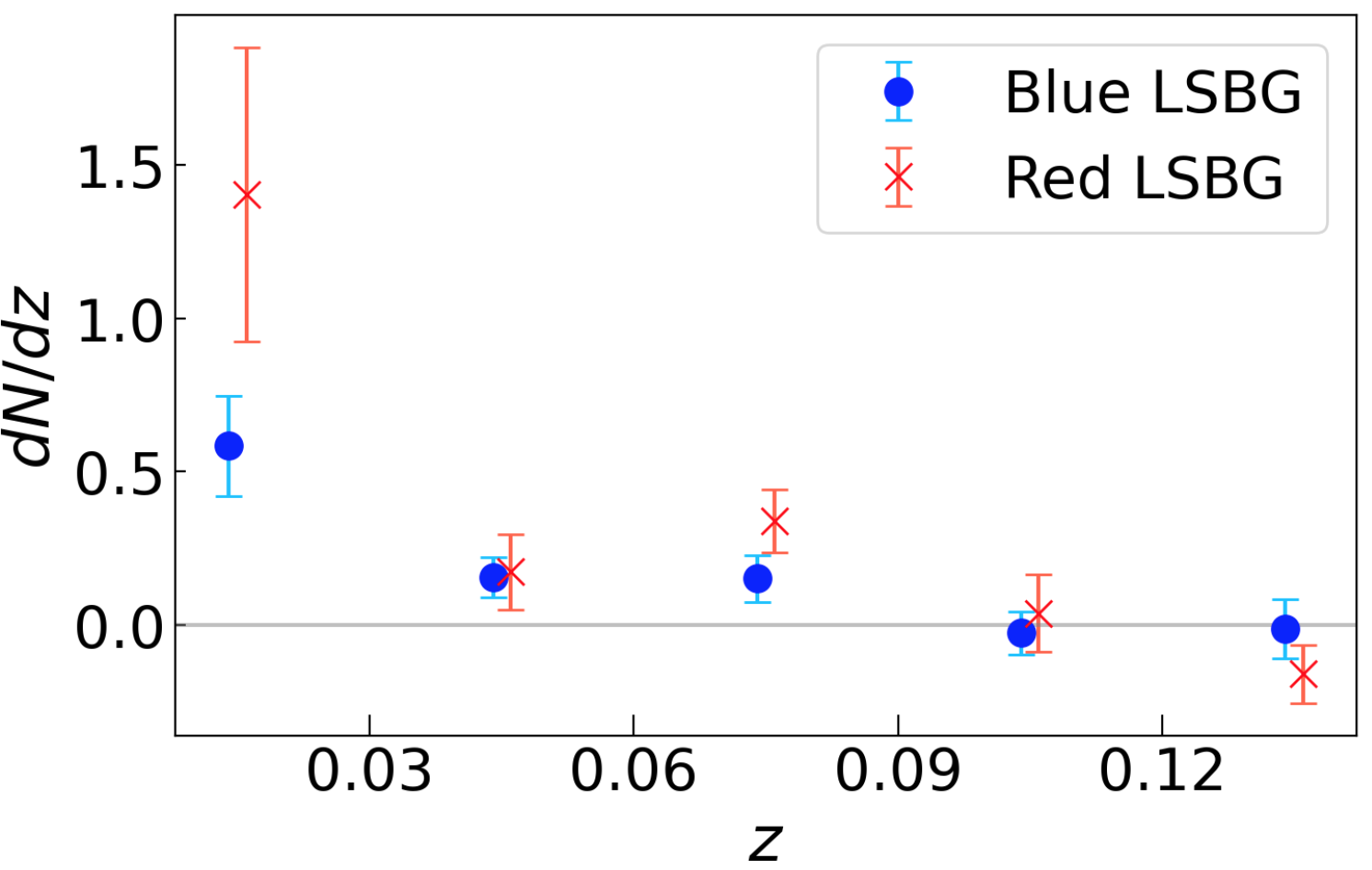}
 \end{center}
 \caption{The $dN/dz$ of blue (blue colored) and red (red colored) HSC-LSBGs. 
 Note that error bars present 2-$\sigma$ level of an uncertainty by considering error estimation of the Jackknife subsampling for the angular cross-correlation of HSC-LSBGs with NSA samples.
 }
\label{fig:lsbg_dndz}
\end{figure}

\ajnold{The rest of the subsection is devoted to describing the covariance of the estimation, which will be used to draw the galaxy redshift randomly from the $dN/dz$ distribution.
We estimate the covariance of $w(z,\theta)$ using the Jackknife method \citep{Scranton-Johnston:2002} as 
\begin{equation}
    C^{m}_{ij} = \frac{M-1}{M}\sum^M_{k=1}\left[ w_{ik}^m - \hat{w}_{i}^m \right] 
             \left[ w_{jk}^m - \hat{w}_{j}^m \right],
    \label{eq:cov}
\end{equation}
where $w_{ik}^m$ is the angular correlation function for the $k-$th Jackknife sub-sample in the $i$-th angular and $m$-th redshift bins. $M$ is the number of the Jackknife sub-samples, and we take $M=100$ which divides the entire region into $\sim 2~$deg$^2$ patches.}
$\hat{w}$ is the averaged correlation function over all jackknife sub-samples,
\begin{equation}
    \hat{w}_{i}^m = \frac{1}{M}\sum^M_{k=1}w_{ik}^m.
\end{equation}
\ajnold{Now the $dN/dz$ at the $m$-th redshift bin can be obtained by integrating $\hat{w}^m_i$ along the angular scale, and its amplitude can be considered as a parameter ${\cal A}_m$.
Therefore, the $dN/dz$ distribution can be stochastically determined with the likelihood function of ${\cal A}$, which is assumed to be Gaussian, possibly with a physically reasonable prior $P({\cal A}<0)=0$ or $1$ otherwise. The variance of the Gaussian likelihood function can be derived as
\begin{equation}
    \hat{\sigma}^2_m \equiv \sum_{i,j}
    \left.\left( \frac{\partial \bar{w}^m}{\partial w^m}\right)\right|_{\theta=\theta_i} 
    C^m_{ij}
    \left.\left( \frac{\partial \bar{w}^m}{\partial w^m}\right)\right|_{\theta=\theta_j},
\end{equation}
\ajnold{where $\bar{w}^m$ can be computed using Equation~\ref{eq:norm_w}.}
}
Figure~\ref{fig:lsbg_dndz} represents the $dN/dz$ distribution measurement. The errors are measured using the Jackknife resampling. 
\ajnold{For the clustering analysis described in this section, we use a publicly available code, }\texttt{treecorr} \citep{2015ascl.soft08007J}.


\subsection{Evaluation of statistical uncertainty}
\label{ssec:uncert}
\reDold{The upper limit on the cross section in the composite analysis can be affected by the $dN/dz$ measurement error and the halo property estimation, which is specified by the halo mass and concentration parameter as described in Section~\ref{ssec:composite}.}
We evaluate 
\omold{the coverage of the upper limits by} using
500 Monte 
\ajnold{Carlo}
simulations.
First, for the halo mass, we evaluate its uncertainty as $\Delta \log {M_{\rm halo}}=0.8$ at 1-$\sigma$ Gaussian error by computing the scatter of stellar-to-halo-mass conversion.
In addition, we adopt a concentration parameter error of $\Delta \log {c}$ of $0.1$ at 1-$\sigma$ Gaussian error \citep{Dutton+:2014}.
Accordingly, the total uncertainty for halo properties results in $J$-factor uncertainties of $\sim$0.9 dex at 1-$\sigma$ error.
\Dold{Moreover, we randomly assign the distance to galaxies according to this distribution; therefore, the negative values of the 
\ajnold{amplitude are physically unreasonable. Therefore, as described in the previous section, we obtain the posterior function of ${\cal A}_m$
as $P({\cal A}_m|{\cal D})=P({\cal D}|{\cal A}_m)|P({\cal A}_m)$.
}
We finally apply linear interpolation between each center of redshift bin to the posterior distribution.}
We take a conservative limit of the minimum redshift of the sample corresponding to 25 Mpc, which is the minimum distance among the HSC-LSBG samples with precisely measured distance.

\begin{figure}
 \begin{center}
  \includegraphics[width=9cm]{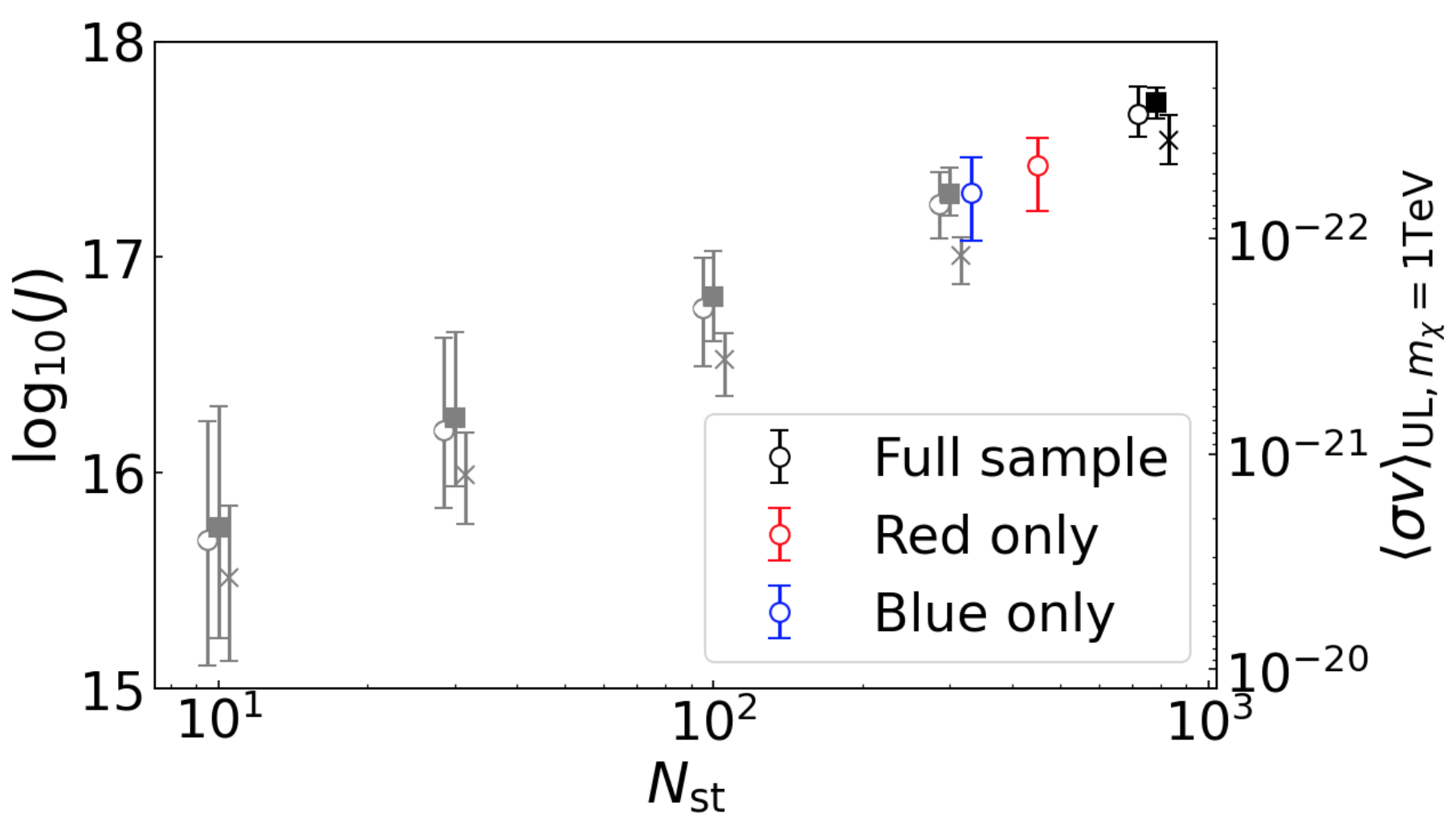}
 \end{center}
 \caption{
    Total J-factor values of HSC-LSBG summed over $N_{\rm st}$ samples taken randomly from 781 samples.
    The error bar with filled square includes $dN/dz$ measurement uncertainty, cross symbol includes halo property uncertainty and circle symbol includes both. Each error bar shows 95\% confidence region based on 500 Monte Carlo simulations.
    The right axis shows $\svul$ \Dold{for $b\bar{b}$ channel} at DM mass of 1~TeV with 95\% C.L. corresponding to J-factor value in the left axis.
    \Dold{The red and blue circles with errorbars show total J-factor values with the both uncertainties for all red and blue LSBGs, respectively.}
    }
 \label{fig:sumJ}
\end{figure}

In Figure~\ref{fig:sumJ}, we show the total J-factor values as a function of the number of stacked objects, $N_{\rm st}$. 
In the order of square, cross and circle symbols, the error-bars plot the total values including $dN/dz$ measurement uncertainty, halo property and both, respectively.
For comparison with other studies, 
we display $\svul$ \omold{at the 95\%~C.L.} for a DM mass of 1~TeV in the right axis, 
\omold{and their corresponding} J-factor value on the left axis.
Note that when converting the J-factor to $\svul$, we apply a mean flux of our UGRB sky.
The upper limit is affected by both the fluctuation and target's $J$-factor value, however we find that, even in the lower energy regime, the scatter by the fluctuation is smaller ($\sim$0.4 dex at 2-$\sigma$ level) than the total halo property and $dN/dz$ measurement uncertainty.

\section{Discussion}
\label{sec:discussion}

\subsection{Correlation between neighbors}
\label{ssec:cov}
We performed the composite analysis in Section~\ref{ssec:composite} assuming that the likelihood functions of individual objects are independent of each other. 
Such correlations are expected when the number density of the sample is high because the PSF of \textit{Fermi}-LAT is $\sim 1^{\circ}$. In this subsection, we demonstrate that the correlation between data at different points can be negligible.

\begin{figure}
 \begin{center}
   \includegraphics[width=8.5cm]{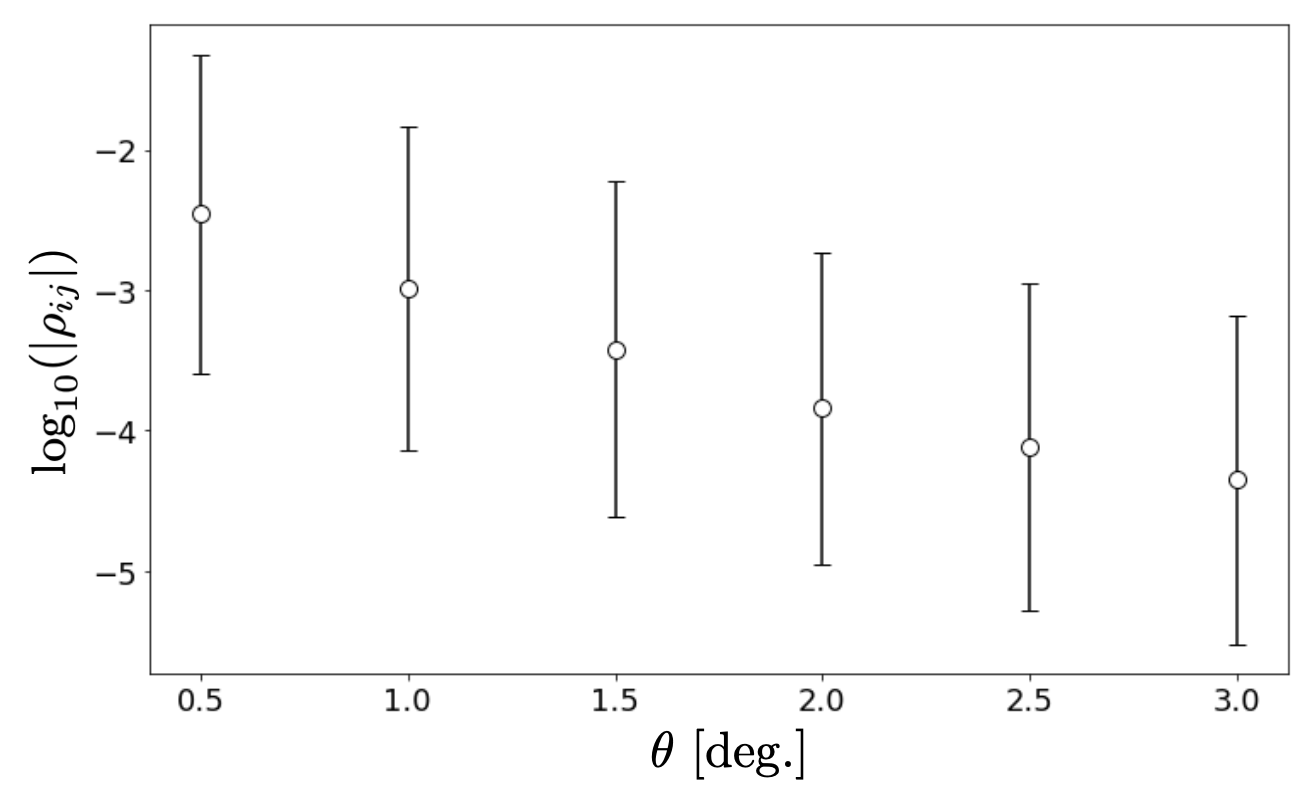}
 \end{center}
 \caption{
    \Dold{The correlation between two amplitude parameters for each paired objects in the overall energy ranges, as a function of separation angles. The error bars represent 1-$\sigma$ errors of the correlations \ajnold{derived from} 60 \ajnold{independent} pairs in each angular bin.}
    }
\label{fig:cov}
\end{figure}

\begin{figure}
 \begin{center}
 \includegraphics[width=8.5cm]{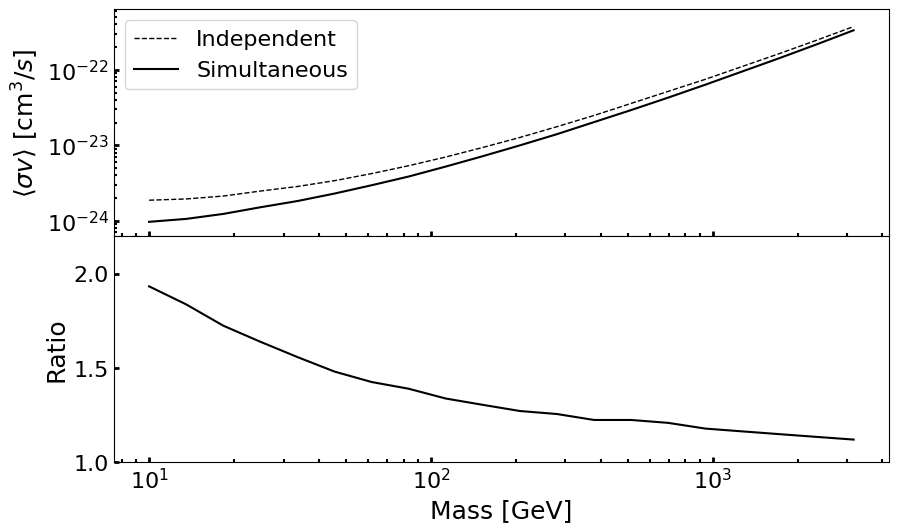}
 \end{center}
 \caption{
    (Top) The difference between $\svul$ \Dold{for the $b\bar{b}$ channel} in the composite analysis with all LSBGs in the simultaneous approach (solid line) and the one with assumption of flux likelihood profiles of all LSBGs being independent of each other (dotted line) as a function of DM mass. (Bottom) The ratio of the upper limits in the two cases.
 }   
 \label{fig:corr_uncorr}
\end{figure}

In the HSC-\textit{Fermi} sky coverage, we select 10 independent patches with the size of $10\times10$ deg$^2$. From each patch, we randomly select 60 pairs of points with separations of 0.5$^{\circ}$ to 3$^{\circ}$. To evaluate the correlation between the putative flux amplitudes of the paired objects, we perform the joint likelihood analysis for the pairs, which simultaneously optimizes the fluxes of the paired objects.
The putative flux of an object is defined by,
\begin{equation}
    \frac{d\Phi}{dE} = \alpha \left( \frac{E}{1000 [\rm MeV]} \right)^{\beta}.
    \label{eq:put_flux}
\end{equation}
\Dold{We fit the amplitude parameters to the UGRB flux and measure the covariance matrix using the \tt{fit} and \tt{sed} methods in \tt{fermipy}.}
Figure~\ref{fig:cov} shows the absolute value of the correlation between two amplitude parameters \Dold{in the overall energy ranges}, as the function of the separation. The error-bars are computed from the 60 independent pairs that reflects the fluctuations of the residual gamma-ray flux. The cross-covariance is normalized by the diagonal terms, i.e., $\rho_{ij} \equiv {\rm cov}(\alpha_i, \alpha_j)/\sigma_i \sigma_j$. Although we expect a strong correlation on scales smaller than 1 deg, 
the correlation at the smallest separation, which corresponds to the HSC-LSBG mean separation, 
is less than 0.1 deg at 1-$\sigma$ level. We note that all the cross correlation \omold{values are} negative at all scales, which is the consequence of the conservation of the total flux.
%
%
For further validation, 
we perform a composite likelihood analysis in which we obtain the likelihood profiles for the putative fluxes of all samples
within a single LAT data patch simultaneously. 
Figure~\ref{fig:corr_uncorr} compares the $\svul$ constraints with this simultaneous approach (`simultaneous' case) with the one obtained based on the assumption that all objects are independent of each other (`independent' case).
We emphasize that the `independent' case gives a weaker constraint on $\svul$
%
%
than the `simultaneous' case. 
This is because the 
total flux conservation is imposed, which results in \omold{a} larger putative flux amplitude \omold{for} the `independent' case. 
%
%
Note that in this calculation, we set all objects' J-factor to $10^{14.5}$~$\rm GeV^2/cm^5$, and choose a specific $dN/dz$. \omold{This explains why} 
the amplitudes of $\svul$ in Figure~\ref{fig:corr_uncorr} do not correspond to the ones in Figure~\ref{fig:sumJ}.

We conclude that the correlation between neighboring points is less than $10\%$, on scales \omold{of order} 0.5 deg. Even if we ignore this correlation, which is computationally much less expensive, we 
\omold{would} obtain 
conservative constraints on \st{the} $\svul$.

\subsection{The power of stacking}
\label{ssec:stacking}
\Dold{In this subsection, we discuss a scaling relation of the statistical power on $\svul$ as a function of the number of \omold{objects included in the} stacking \omold{procedure}. We 
\ajnold{first analytically show that} $\svul$ is proportional to the inverse of $N_{\rm st}$ 
\ajnold{in the limit of }
zero observed photon counts.}
\Dold{We 
then also demonstrate that this scaling relation converges to $N_{\rm st}$ 
\ajnold{independent of }
the DM mass if the $N_{\rm st}$ is sufficiently large \ajnold{by using } HSC-LSBG catalog and Fermi observations.}

\ajnold{Let us first revisit}
the Poisson likelihood for a single LSBG in the UGRB. 
The likelihood function for the model parameters is given by the Poisson distribution,
\begin{equation}
    {\cal L} = \prod_{i} \frac{\lambda_{i}^{n_{i}} {\rm e}^{-\lambda_{i}}}{n_{i}!},
\end{equation}
where the index $i$ runs over all energy bins and pixels and $n_i$ is the observed photon counts;
$\lambda_i$ is expected photon counts for model fluxes, which is decomposed into the Galactic foreground, isotropic background and resolved point-source model fluxes as well as the flux by the DM annihilation for the single LSBG. 
A UGRB sky is derived from the modeling of all the \gray sources except for the LSBG flux as described in Section~\ref{sssec:measured_flux}. Therefore, 
\omold{if we fix} the model parameters $\boldsymbol{\Theta}$ of the background sky,
%
%
%
\omold{then} $\lambda_i$ depends only on the 
parameter $\sv$.
We denote $\lambda_{i} = \lambda_{i}^{\rm others}(\boldsymbol{\Theta}) + \lambda_{i}^{T}(\sv)$, where $\lambda_{i}^{\rm others}$ is the total model flux of all sources without LSBG and $\lambda_{i}^{T}$ is the model flux of LSBG.
%
%
For simplicity, we consider that all LSBGs have the same J-factor, which means that their model fluxes are equal.
In addition, we consider the fact that, in energy regimes higher than $\sim$30~GeV, the LAT hardly detects photons.
Given that there is no photon count ($n_i=0$) in such energy regimes and in all pixels, we expect $\lambda_i^{\rm others}(\mu)=0$ and then $\lnl = -\sum_i \lambda_i^T(\sv)$.
Consequently, we obtain the composite likelihood with $N_{\rm st}$ objects using Equation~\ref{eq:stackL},
\begin{equation}
    \Delta \lnl_{\rm st} = -N_{\rm st}\sum_i \lambda_i^T(\sv),
\end{equation}
where $N_{\rm st}$ is the number of objects in the composite analysis.
Thus, in our 
\omold{approach} the upper limit is proportional to $1/N_{\rm st}$ 
\omold{since} $\lambda_i^T(\sv) \propto \sv$. 

\begin{figure}
 \begin{center}
   \includegraphics[width=8.5cm]{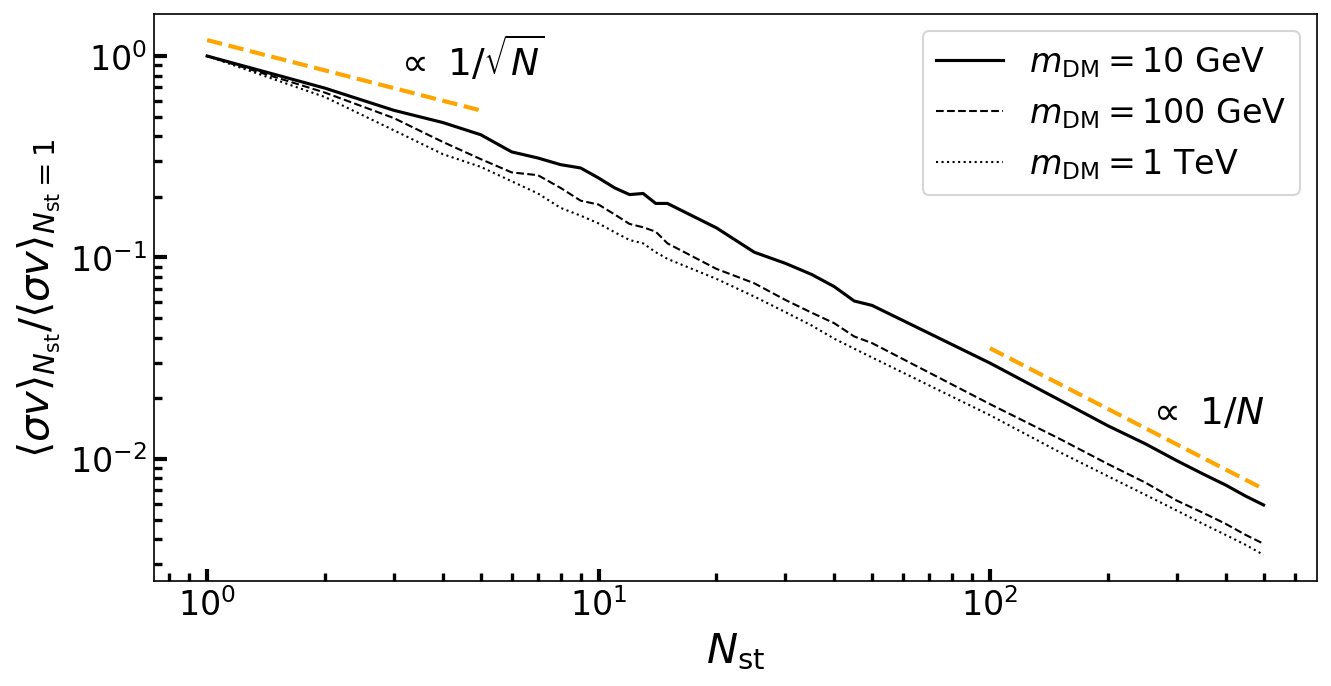}
 \end{center}
 \caption{
    Scaling relation of $\svul$ with a number of objects $N_{\rm st}$ in the composite analysis. 
    The vertical axis is the ratio of the upper limit with $N_{\rm st}$ objects to the one with a single object.
    The black solid, dashed and dotted lines correspond to the ratio for DM mass of 10~GeV, 100~GeV and 1~TeV, respectively.
    Scaling with $1/N_{\rm st}$ and $1/\sqrt{N_{\rm st}}$ are shown in orange dashed lines. \Dold{We set the annihilation channel to $b\bar{b}$.}
    }
 \label{fig:scaling}
\end{figure}

\Dold{Further, we explore the scaling relation using the HSC-LSBG catalog and Fermi data.
We calculate the 
\ajnold{constraint on $\sv$ from randomly drawn $N_{\rm st}$ objects}
from the HSC-LSBG \ajnold{sample.} 
\ajnold{Because here we aim to isolate the effect of number of stacking from the individual halo properties, we assume the selected sample has a same $J$ factor, but using the actual background \gray fluxes at that position.}
We repeat this calculation 500 times for each $N_{\rm st}$ sampling.}
In Figure~\ref{fig:scaling}, we show the median value of the ratio of $\sv_{\rm UL}$ with $N_{\rm st}$ objects to the one with a single object for DM mass of 10~GeV, 100~GeV and 1~TeV.
With $N_{\rm st}$ larger than $\sim$30, the upper limits scale with the inverse of $N_{\rm st}$ for all mass ranges.
\Dold{We have discussed this relation in our previous paper~\citep{2020JCAP...01..059H}, which 
\ajnold{predicted the }
relation 
\ajnold{based only on }
8 HSC LSBGs 
. \ajnold{In this paper, we confirm that prediction of scaling with $N_{\rm st}$, but furthermore we find, thanks to the large number of LSBG sample, that the relation is independent of DM mass.}
}
\ajnold{This is because most of the constraints are dominated by the objects having significantly low background fluxes. Such objects can be found with a certain probability, say $\sim 10\%$ of the entire sample. Therefore, when the number of stacking objects is small, e.g. $N_{\rm st} <10$, the probability of having strongly constrained objects is less than $10\%$, and the constraining power 
\omold{does not scale} with $1/N_{\rm st}$. However, the probability of having such objects may increase and converge to $10\%$. as we increase the number of stacking. Therefore, the statistical power of stacking, dominated by the objects with almost zero-background fluxes scales with $1/N_{\rm st}$.}
For DM mass of 100~GeV and 1~TeV, this scaling relation is seen, even in smaller $N_{\rm st}$.
This behavior is reasonable, considering the photon-count statistics in a high energy regime.
The annihilation process with more massive DM particles can produce higher energy photons, thus probing the DM annihilation for more massive DM is affected by photon-count statistics in high energy regimes.

\section{Conclusions} 
\label{sec:summary}

In this study, we proposed a \ajnold{new} method to
\Dold{constrain the DM annihilation cross-section with} potential \gray sources with unknown redshifts \Dold{and the observed \gray background}.
The 
\omold{crucial} point is that we do not need to 
\ajnold{know}
the individual distance to the \gray sources. The overall redshift distribution is sufficient to constrain the DM annihilation cross section.
By applying the redshift clustering method, we obtained a redshift distribution of 
\omold{the whole population,} 
instead of \omold{the} redshifts of individual objects. 
\omold{We then} randomly assigned the distance to each LSBG according to this distribution.
To validate the proposed method, we measured $dN/dz$ of full samples of the HSC-LSBG catalog including $\sim$800 objects and constrained the DM annihilation cross-section via a composite likelihood analysis using all LSBGs given distances \omold{and using the eight years of \textit{Fermi}-LAT observations.} 

In the $dN/dz$ measurement for the LSBG samples, we employed the NSA catalog as spec-z samples in overlapped regions with the HSC region.
The uncertainty in the $dN/dz$ 
\omold{was} evaluated with 
Monte Carlo simulations. 
\reDold{We found that in the composite analysis with $\sim$1000 objects,
uncertainties of the halo property estimation and $dN/dz$ measurement are quite small and not significantly affected to the upper limit on the cross section.}
Therefore, using $dN/dz$ distribution of overall target samples instead of the individual sample's redshifts is a valid and robust approach for the DM signal search.

To reduce the computational cost, we 
\omold{assumed} that all \omold{the} 
\omold{spectral} parameters \omold{of the} different LSBGs are independent of each other.
The number density of the LSBGs ($\sim 0.5^{\circ}$) was comparable to the LAT PSF scale, which may lead to a correlation between 
\omold{spectral} parameters for neighboring LSBGs, and thus may break the assumption.
We validated 
\omold{this conjecture} in two ways. First, we computed \omold{the} covariance matrix between \omold{the} two 
\omold{normalizations} for each paired neighbor and 
\omold{checked} that off-diagonals \omold{terms} of the correlation matrix is smaller than 0.1 at the separation angle $0.5^{\circ}$.
Second, fixing the J-factor and distance of each object, we performed a composite likelihood analysis in which we obtained the likelihood profiles for putative fluxes of all samples in each LAT patch simultaneously. By comparing $\svul$,
we found that the constraints differ at most \omold{by a} factor of 2. Moreover, if we 
\omold{assume that} the data \omold{is independent}, 
\omold{then} the constraints become more conservative due to the relaxation of the total flux conservation condition. 

Finally, we found the scaling relation of $\svul$ with the number of objects $N_{\rm st}$ in the composite likelihood analysis. \Dold{First, under an assumption of zero observed photon, we discussed the relation analytically and} found that $\svul$ is proportional to $1/N_{\rm st}$.
\Dold{Furthermore, using the real HSC-LSBG catalog and the LAT data,} we computed the scaling for DM mass of 10~GeV, 100~GeV and 1~TeV and showed that for all DM masses the upper limit scales with $1/N_{\rm st}$ using $N_{\rm st} \gtrsim 30$. 
This is a significant effect which is the consequence of the Poisson statistics \Dold{due to rare photons detected by the LAT} and is different from the Gaussian statistics in which the scaling obeys  $1/\sqrt{N_{\rm st}}$.

In future imaging surveys with next-generation telescopes such as the LSST, a 
\omold{large number} of LSBGs will be detected because of wider sky coverage and better sensitivity.
For example, LSST has a sky coverage of $\sim$20,000~deg$^2$ and reaches the depth of $\sim$27.5~${\rm mag/arcsec^2}$ in 
$i$ band and consequently has the potential to discover ${\cal O}(10^5)$ objects.
Although, for the reference samples we need spec-z or high-precision photo-z samples located in the local universe,  
it is expected \omold{that these will} decrease \omold{the} uncertainties 
\omold{in the} estimate of $\svul$ due to \omold{the} $dN/dz$ measurement and \omold{the} halo properties.
Moreover, by increasing the statistics, we can 
\omold{perform} a more detailed $dN/dz$ measurement, particularly \omold{in} a redshift range corresponding to \omold{a} distance of less than 25 Mpc, which is the minimum distance adopted \omold{here} for \omold{a} conservative J-factor estimate in our likelihood procedure. 
\omold{We expect that in the future, }this will give a more stringent constraint on the DM cross-section.

\begin{acknowledgments}
This work was supported in part by World Premier International Research Center Initiative, MEXT, Japan, and JSPS KAKENHI Grant No.~19H00677, 20H05850, 20H05855, 20J11682 and 21H05454.
\end{acknowledgments}


\bibliographystyle{unsrt}
\bibliography{bibdata}

\end{document}